\newcommand{\eq}{\begin{equation}}
\newcommand{\en}{\end{equation}}

\documentclass[12pt,manuscript]{aastex}

\usepackage{amsmath}
 
\input epsf              
 
\begin{document}

\title{ Characterization of the near-Earth Asteroid 2002~NY$_{40}$\altaffilmark{1}}

\author{Lewis C. Roberts, Jr.,
Doyle T. Hall,
John V. Lambert,
John L. Africano\altaffilmark{2},
Keith T. Knox,
Jacob K. Barros,
Kris M. Hamada,
Dennis Liang,
Paul F. Sydney}
\affil{The Boeing Company, 535 Lipoa Pkwy, Suite 200, Kihei, HI 96753}

\author{Paul W. Kervin}
\affil{Detachment 15, Air Force Research Laboratory, Directed Energy Directorate, 535 Lipoa Pkwy, Suite 200, Kihei, HI 96753}

\altaffiltext{1}{Based on data from the Maui Space Surveillance System, which is operated by Detachment 15 of the U.S. Air Force Research Laboratory's Directed Energy Directorate.}

\altaffiltext{2}{Deceased}

Pages: 24

Tables: 0

Figures: 4

\newpage

Proposed Running Head : Characterization of 2002~NY$_{40}$
 
Editorial Correspondence to:

Lewis Roberts,
Boeing,
535 Lipoa Pkwy, Suite 200,
Kihei HI 96790.

Phone: 808-874-1653 Fax: 808-874-1600

Email: lewis.c.roberts@boeing.com

\newpage


\section*{Abstract}

In August 2002, the near-Earth asteroid 2002 NY$_{40}$, made its closest approach to the Earth.  This provided an opportunity to study a near-Earth asteroid with a variety of instruments.  Several of the telescopes at the Maui Space Surveillance System were trained at the asteroid and collected adaptive optics images, photometry and spectroscopy.  Analysis of the imagery reveals the asteroid is triangular shaped with significant self-shadowing.  The photometry reveals a 20-hour period and the spectroscopy shows that the asteroid is a Q-type.

Key Words: asteroids; near-Earth objects; spectroscopy; adaptive optics

\newpage


\section{Introduction}

2002~NY$_{40}$ was discovered by the LINEAR system in New Mexico on 14 July, 2002 \citep{tichy2002}.  Just over a month later, on 18 August, it came within 0.0036 AU of the Earth.  During this approach its visual magnitude reached as high as 8th magnitude.  Soon after its closest approach, the phase angle changed and the object's brightness quickly dropped to below 18th magnitude.  The approach provided an ideal opportunity for observation; as it was bright and had a large angular diameter.  The brightness allowed the use of the AEOS Adaptive Optics  (AO) system to image the asteroid.  Before and during the approach we were also able to collect photometry and low-resolution visible spectroscopy.  The combination of these data sets allow us to constrain physical properties of the asteroid as well as provide a limited understanding of its morphology.

We acquired data with sensors on two telescopes at the Maui Space Surveillance System (MSSS) located at the summit of Haleakala.  These were images with the AO system on the 3.6 m AEOS telescope and spectra with the Spica spectrograph on the 1.6 m telescope. In addition  we collected photometry data with the 0.36 m Raven telescope at the Remote Maui Experiment (RME) site at the base of Haleakala.


\section{Photometry}

\subsection{Data Collection}

The Raven telescope \citep{sydney2000} was designed as an autonomous, affordable, and robust system, capable of delivering highly accurate astrometry and photometry.  The system was designed to be low-cost, which allows for multiple versions expanding the  data collection capabilities.  

These observations used the Raven telescope at the Remote Maui Experiment (RME) site at the base on Haleakala in Kihei HI.  At the time of the observations, the imaging sensor was a 512 pixel by 512 pixel thinned back-illuminated Apogee AP-7 CCD camera.  The AP-7 camera uses the SITe SIA502AB CCD chip with 24 mm pixels, 16 bit A/D at 30 kHz, ~14 photoelectrons noise with a gain of 5.5 photoelectrons per ADU, and a responsivity from 3000--10,000 \AA~with a peak responsivity of ~85\%
 between 6000--8000~\AA.  The CCD camera is complemented with a two stage thermoelectric cooler running at a nominal -25$^{\circ}$C.  The camera is unfiltered but has a detection spectral band similar, but not identical, to V-band. In this configuration, the telescope has a 38\arcmin~square field of view that allows it to track objects traveling at rates up 45\arcsec~per second.  This is not normally of great utility for astronomical observations, but it was ideal for observations of the fast moving 2002~NY$_{40}$, allowing for long exposure images without smearing. The RME-Raven telescope acquired photometric measurements 2002 August 13--18.   The exposure time was adjusted to keep a constant signal level on the asteroid ranging from 15s to 60s as the object got fainter. 

Due the asteroid's orbital motion, 2002~NY$_{40}$ was moving much faster than sidereal rate.  So the background stars quickly changed over the course of several frames.  This made it impossible to compute differential photometry against a common set of stars.  Instead the differential magnitude was computed against whichever stars were in the field of view.  Asteroid and field-star signals are extracted from RME-Raven telescope images using a photometry reduction algorithm similar to that of \citet{stetson1987}.  Correlating the open-filter magnitudes of observed field-stars to their cataloged V-band magnitudes provides a rough estimate of the RME-Raven instrumental zero-point.  However, this open-filter to V-band magnitude conversion process results in a $\pm$0.25 magnitude absolute calibration uncertainty.  Because of this uncertainty, the RME-Raven photometric data have not been used to determine the asteroid's albedo or size, and have only been used to constrain the rotation period and modulation amplitude.
 
\subsection{Data Analysis}

\citet{pravec2005} carried out a more complete photometric analysis using a gread deal of data from a number of observatories including the first three of the data sets presented here. This analysis found the asteroid had a main period of 19.98$\pm$0.01 h with a secondary period of 18.43$\pm$0.01 h, indicating that the asteroid is tumbling with two principal axes.  They also were able to determine that the asteroid has large non-convex features.  

Using the period of \citet{pravec2005}, we fit a 2-parameter IAU asteroid photometric model \citep{bowell1989} augmented with a simple sinusoidal variation to our data.  Figure \ref{lightcurve} shows the measured and model light-curves for 2002~NY$_{40}$.  Both the measurements and the model curves in Fig. \ref{lightcurve} have been converted to reduced magnitudes, i.e. adjusted to remove 1/R$^2$ variations by normalizing the Sun-asteroid and observer-asteroid distances to 1~AU (no phase angle corrections have been applied).  The dotted line in Figure \ref{lightcurve} shows the expected variation using model parameters $H$\,$=$\,19.0 and $G$\,=\,0.15.  This model generally follows the long-term photometric variations caused by the changing solar phase angle, but clearly does not account for the observed modulations.  The dashed line in Fig. \ref{lightcurve} shows the same 2-parameter model but augmented with a best-fit sinusoidal variation with a peak-to-peak period of 19.98~hour \citep{pravec2005} and peak-to-peak amplitude of 0.96 magnitudes. We include the light curve in Fig. \ref{lightcurve} only in order to provide context for the adaptive optics imagery and spectroscopic observations that are discussed in the following sections. The times of those observations are indicated with vertical lines. 


\textbf{Place Figure 1 here. }


\section{Images}\label{sec_images}

Several near-Earth asteroids have been imaged by radar \citep{hudson2003,benner2002}, and several main belt asteroids have been imaged with speckle interferometry \citep{mccarthy1994,ragazzoni2000} and AO \citep{saintpe1993,drummond1998,hestroffer2002, baliunas2003}.  For the most part near-Earth asteroids are too faint for speckle interferometry and AO systems.  2002~NY$_{40}$ came very close to the Earth and presented a unique opportunity for optical observations.  In addition to the observations reported here, the object was imaged with the AO system on the 3.5 William Herschel telescope in H-band\footnote{http://www.ing.iac.es/PR/press/ing32002.html}.  \citet{howell2003} reports that 2002~NY$_{40}$ was also observed with radar, so the object will have imagery in the visible, near-infrared and radar.  These observations occurred at different times and this temporal coverage should provide a more complete understanding of the body's morphology as it rotates.

\subsection{Data Collection}

The images were taken with the AO system \citep{roberts2002} on the 3.6 m AEOS telescope.  $I$-band images were acquired from 5:24--5:35 on 2002 August 18 UT while the asteroid was at a zenith distance of $20^\circ$. This was the asteroid's closest approach and before this the asteroid was expected to be unresolved.  As shown in Fig. \ref{lightcurve}, the asteroid's brightness quickly faded as the phase angle changed, so the AO correction was best at the beginning of the observations.  By the end of the observations, the asteroid had faded enough that the AO system was no longer providing useful data. 

The exposure time was varied from 0.998-9.998 seconds in an attempt to find the optimal balance between collecting enough signal and not introducing too much smear to the image because of uncompensated atmospheric turbulence.   During the observations the seeing ranged ranged from  1.2-2.0\arcsec, as reported by the Day Night Seeing Monitor at the observatory \citep{bradley2006}.  This is worse than average, and in combination with the relatively dim object for the AO system, produced data with below average image quality. Frames were taken as long as the target was bright enough to provide enough signal to enable the AO system to operate.

\subsection{Data Analysis}

After the observations, the individual frames were debiased, dark subtracted and flat fielded using standard techniques.  Then the images were processed using a Richardson-Lucy blind deconvolution technique.  The Richardson-Lucy algorithm \citep{richardson1972,lucy1974} is an iterative image restoration method that assumes a Poisson-distributed noise model to deconvolve an image using a known point-spread function (PSF).  For those cases in which the PSF is unknown, a method, called blind deconvolution, was suggested by \citet{ayers1988} to iteratively estimate both the object and the PSF.

Several authors have combined the two iterative techniques together to form iterative techniques that alternate between estimating the object and the point-spread function \citep{paxman1992, tsumuraya1994, fish1995}. The implementation of Richardson-Lucy blind deconvolution, that was used for this experiment \citep{gerwe1999} incorporates multiple input frames and can process frames in which the images are partially outside of the field-of-view.  It alternates between estimating the object and the point-spread function.   Finally, it can perform super resolution, \textit{i.e.} it can estimate the object on a finer pixel grid than the input image frames were taken.

In the processing of the images in this paper, the original $512\times512$ input images were cropped to $256\times256$ and then estimated on a $512\times512$ output pixel grid, yielding a potential super resolution of a factor of two.  Each output frame used 5 input frames to perform the estimation of the object. The iterations were stopped after 500 iterations on each of the object and the individual PSFs. The various images were examined and it was determined that the data frames with exposure times of 0.998 s produced the best images.  These images were collected from 5:24-5:25 UT.  There were 18 frames with this exposure time and three independent images were produced from these images.  These are shown in Fig. \ref{ny40_images}. 

\textbf{Place Figure 2 here. }

Whenever deconvolution is applied to images, caution must be used when analyzing the results as deconvolution can introduce artifacts that can be misinterpreted. The three images are created from independent frames, but produce a fairly similar object.  The same triangular shape was also found in a shift-and-add image and an earlier deconvolution that used a different method \citep{roberts2003}.  We can safely conclude that the asteroid has a roughly triangular shape with significant self shadowing.  The low level details of the images change significantly from image to image and are not credible.  Since these are visible light images, it is quite possible that a portion of the asteroid is in shadow and not appearing in these images.  Also we are imaging a two-dimensional projection of a three-dimensional object.  We get very little knowledge of the depth of the features from these images.  The self-shadowing shown in the images agrees with the determination of \citet{pravec2005} that the asteroid has non-convex features.  

The illumination geometry of the asteroid at the time of the observations is shown in Fig. \ref{shadow}.  The orientation of the image is the same as in Fig. \ref{ny40_images}. The projected vector to the Sun is marked.  From the ephemeris, the Sun-Target-Observer angle is 78.6$^\circ$, so the Sun vector is 11.4$^\circ$ out of the page.  The sphere is a little more than half illuminated. As these are  $I$-band images, they only see the solar illuminated portion of the asteroid. We have no information about the dark portion of the asteroid.

\textbf{Place Figure 3 here. }

The range to the asteroid is given by the JPL Solar System Dynamics ephemeris website\footnote{http://ssd.jpl.nasa.gov/horizons.cgi} as $3.5\times10^{-3} \pm 2.5\times10^{-6}$AU. The angular pixel scale for the raw AO images is 0.0211 $\pm$ 0.0004 arcsec as determined from observations of binary star with well known orbits \citep{roberts2002}. As mentioned above, the images have been sampled onto a grid twice this during the deconvolution process.  Combining this information gives a linear pixel scale on the asteroid of 0.027 km/pixel.  Each sub-image in Figure \ref{ny40_images} has a  1 km vertical line drawn on image.  With a non-symmetric object, it is hard to describe it with a single size, but the portion of the asteroid facing the Earth at the time observations is roughly 800 $\pm$ 100 m across.    

These images present some unique challenges.  AO imaging of asteroids typically occurs when the asteroid is at opposition, eliminating the problem of self-shadowing.  Imaging of near-Earth asteroids is most commonly done with radar  \citep{hudson2003,benner2002}, which again does not have the self-shadowing problem.  In addition the objects are observable over a longer period which allows measurement of multiple sides of the object, allowing for a determination of the 3-dimensional shape.  Our observations were only possible during the closest approach of the asteroid, and thus we were only able to observe one aspect of the object. This limits our ability to fully understand the object.  A full understanding requires additional data during subsequent near-Earth approaches or with other techniques. 

There have been a range of possible sizes for the asteroid ranging from 250-450m \citep{rivkin2003,howell2003,muller2004}.  Our analysis presents yet another size estimate and it is larger than the rest. The different sizes come from different techniques and it is hard to resolve the differences between the techniques without more information or additional observations.  From the images shown here it is obvious that the asteroid deviates significantly from spherical.  It maybe that the different techniques measured the asteroid at different times and represent different projections of the asteroid.  


\section{Spectroscopy}\label{sec_spectroscopy}
    
\subsection{Spica Spectrograph}

The Spica spectrograph \citep{nishimoto2001} was designed for spacecraft observations, in particular the discrimination of classes of objects and the determination of spacecraft health and status \citep{jorgensen2004}. The system is designed around a commercial off-the-shelf Acton Research spectrograph that features a grating turret with three selectable positions.   For these observations we used a relatively broad-range 150 lines/mm grating. The 150 lines/mm grating produces a spectral resolution of 3--5~\AA~for point source observations.

Although Spica possesses the capability to select any central wavelength for a given observation, it typically operates in two modes, a blue-wavelength mode that ranges from 4000--7000~\AA~and a red-wavelength mode that covers 6000--9000~\AA.  The blue mode currently operates with a long pass filter configuration blocking all wavelengths shorter than 4000~\AA, while the red-mode is equipped with an order-separation filter that cuts wavelengths longer than 5500~\AA.  The spectrograph CCD array is 330$\times$1100 pixels with a pixel size of 25~$\mu$m equipped with an on-chip summing capability and cooled by liquid nitrogen. The system has mercury and neon emission lamps for use as wavelength calibration sources.

\subsection{Data Collection} \label{spectra_collection}

Spectral data were taken of 2002~NY$_{40}$ on 2002 August 17 \& 18. On the 17th, blue spectra were taken from 9:10--9:54 UT and red spectra were taken from 10:59--11:25 UT.  Additional red data were taken on the 18th from 8:41--9:17 UT.  A sky background spectrum was taken after every asteroid spectrum.  Each spectrum had an exposure time of 60~s and used the 150 lines/mm grating.  Observations were collected of the solar analog stars: SAO~126133 and SA~107-998.  These stars were chosen so that the airmass matched as closely as possible with the observation airmass of the asteroid.   Also, the magnitudes are similar to the asteroid's predicted magnitude. Spectra of two spectro-photometric standard stars, HIP~107864 and HIP~101516, were collected on both nights.  Finally spectra of the internal calibration sources were recorded.

\subsection{Data Reduction} 

The first step in the data reduction process is to remove cosmic rays from all spectra.  Next the  sky background was removed from each stellar or asteroid spectrum. The spectra were then wavelength calibrated with the spectra of the internal sources. Then the effects of atmospheric extinction, optical transmission, detector quantum efficiency are removed and the intensity is calibrated to ergs~cm$^{-2}$~s$^{-1}$~\AA$^{-1}$. This is done by simultaneously solving the sensitivity function from the observations of the spectro-photometric standards and a fixed site extinction curve.   Finally the asteroid spectrum is divided by the solar analog's spectra.
 
The individual spectra were examined and it was determined that they showed little variation over each night.  This most likely indicates that the asteroid has a fairly homogeneous surface composition. The decision was made to co-add the spectra to improve the signal-to-noise ratio of the data.   We visually compared the 2002~NY$_{40}$ spectra to spectra from asteroids with a wide variety of types.  The match to the Q-type spectra was the best, with Sq and S being second and third closest matches.  Q type asteroids are confined to the near-Earth population and are spectral analogs of  the ordinary chondrite class of meteorites  \citep{lipschutz1989}.  This determination agrees with that of \citet{rivkin2003} who derived the type from infrared spectroscopy and visible spectrophotometry.  The final co-added spectrum from 2003 August 17 is shown in Fig. \ref{spectra}. The offset over-plotted spectra are from the Q-type near-Earth asteroids: 1999 CF$_{9}$ and 2000 AC$_{6}$ \citep{binzel2001}. All three spectra were normalized to 1.0 at a wavelength of 5500 \AA.  The spectra are closely matched, the difference below 5000 \AA~may be an artifact of the use of a fixed site-extinction curve. 
  
\textbf{Place Figure 4 here. }


\section{Conclusion}

Comparing the light curve in Fig.~\ref{lightcurve} and the time of the AO images, shows that the AO system acquired data when the asteroid was brightest and presumably showing its largest side toward the Earth.  This is very lucky; if the images had been acquired when the asteroid was showing its smallest side towards the Earth, the object would have been dim enough that the AO system would not have been able to close the loop.  

The images of the asteroid have shown us that the asteroid is roughly triangularly shaped with significant self-shadowing which is evidence of the non-convex features detected photometrically by \citet{pravec2005}.  The spectroscopy has revealed that the asteroid spectrum is a Q-type asteroid consistent with the analysis of \citep{rivkin2003}.  In addition the spectra taken on different nights are very similar, indicating that the asteroid has a fairly uniform surface composition.

\section*{Acknowledgments}

We thank B. Africano, A. Alday, R. Cortez, K.P. Kremeyer, B.B. Law, and the rest of the staff of the Maui Space Surveillance System for their assistance in taking these data.  This research was funded by AFRL/DE (Contract Number F29601-00-D-0204). We would like to thank Anita Cochran for her useful comments and David Gerwe for supplying the deconvolution code used to produce the images.






\clearpage

\figcaption[]{The 2002~NY$_{40}$ light curve compared to a simple photometric model.  The measured reduced magnitudes of 2002~NY$_{40}$ correspond to the diamonds. The dotted line is the 2-parameter model fit.  The dashed line shows the model augmented with a sinusoidal variation with an peak-to-peak amplitude of 0.96 magnitudes. The vertical line starting at the top of the figure indicates when the adaptive optics images described in \S \ref{sec_images} were taken, while the two bars starting at the bottom of the graph indicate when the spectroscopic observations described in \S  \ref{sec_spectroscopy} were taken.\label{lightcurve}}

\figcaption[]{Deconvolved images of the 2002~NY$_{40}$. The top row of images are deconvolved from five independent raw images.  The bottom image is the average of the three images.  
 The vertical lines are 1 km in length.  The asteroid appears to be a roughly triangular shaped object with significant self shadowing.  The illumination geometry is shown in Fig. \ref{shadow} \label{ny40_images}}

\figcaption[]{The illumination geometry for the asteroid at the time the AO images were taken. For this purpose the asteroid was assumed to be a sphere and is a little more than half illuminated.  The North is indicated with by an arrow labeled N and the Sun vector is to the left at an angle 63 degrees from North. \label{shadow}}

\figcaption[]{The co-added spectra of 2002~NY$_{40}$.   The spectra of 1999 CF$_{9}$ and 2000 AC$_{6}$ \citet{binzel2001} are over plotted in thinner lines.  The spectrum for 1999 CF$_{9}$ is offset above the spectrum of 2002~NY$_{40}$, while the spectrum for 2000 AC$_{6}$ is offset below.\label{spectra}}

\newpage

\begin{figure}[!ht]
\centerline{\epsfysize=4.0in \epsfbox{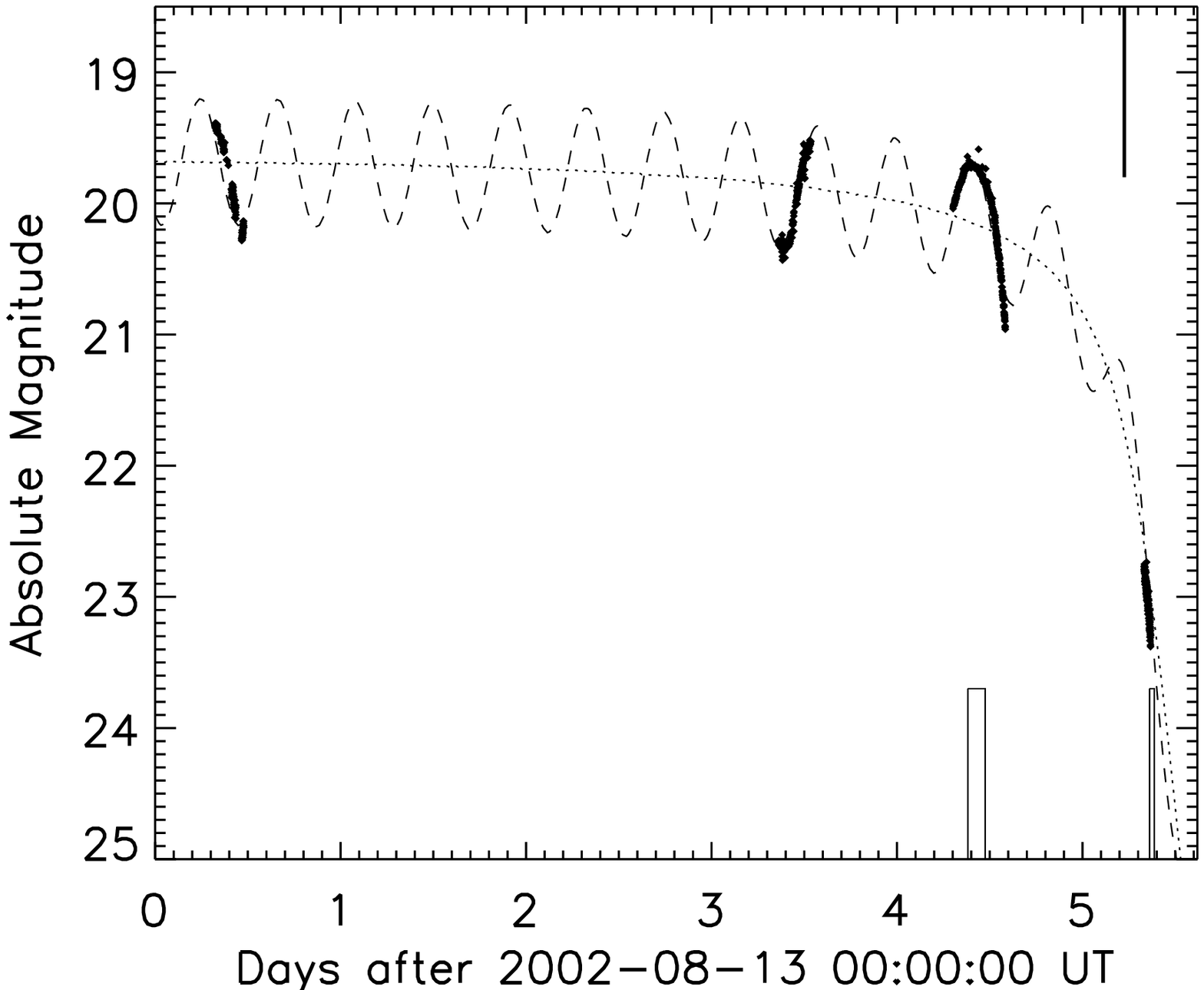}}
\end{figure}

Figure 1. Roberts et al. 

\newpage

\begin{figure}[htbp]
\centerline{
  \epsfysize=2in \epsfbox{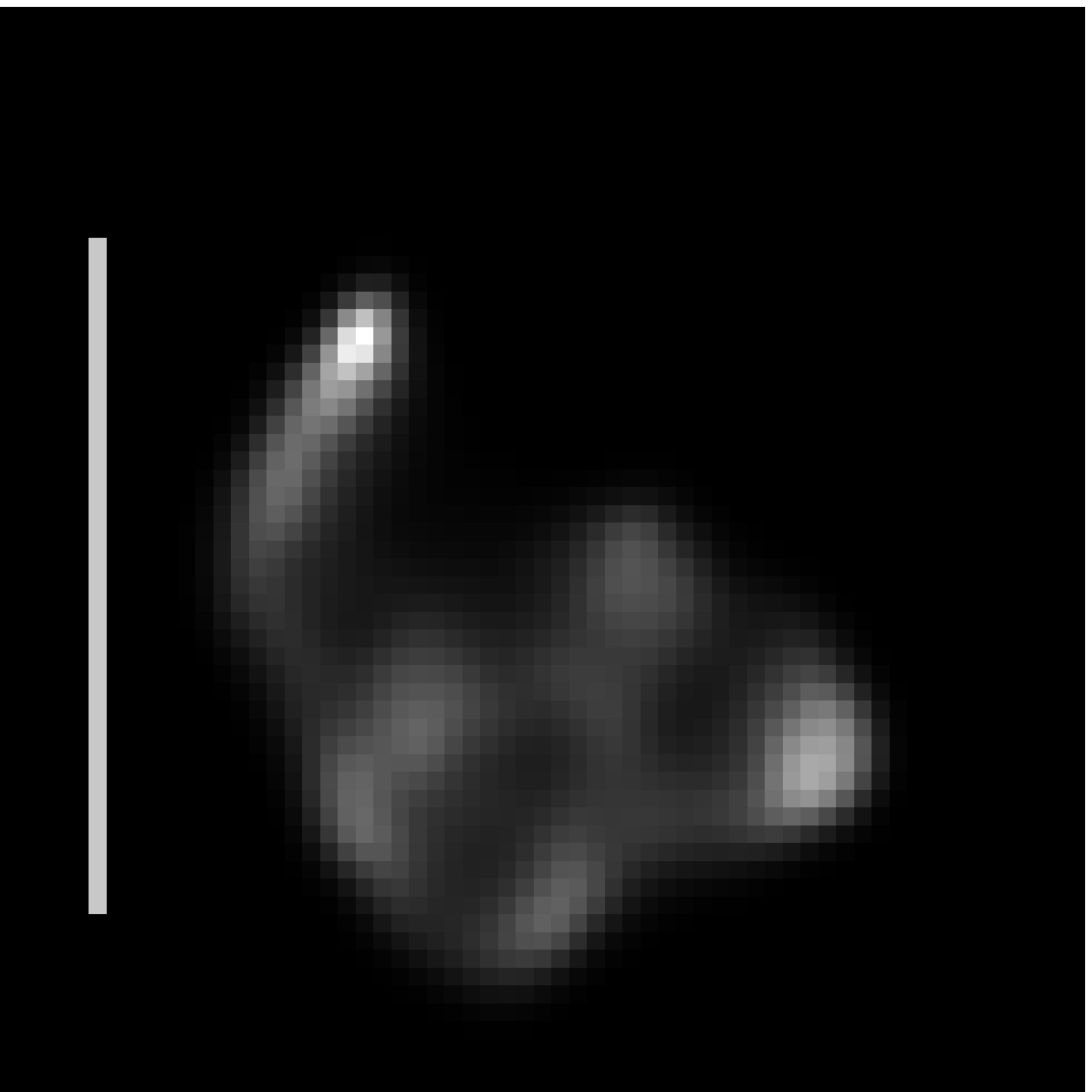}
  \epsfysize=2in \epsfbox{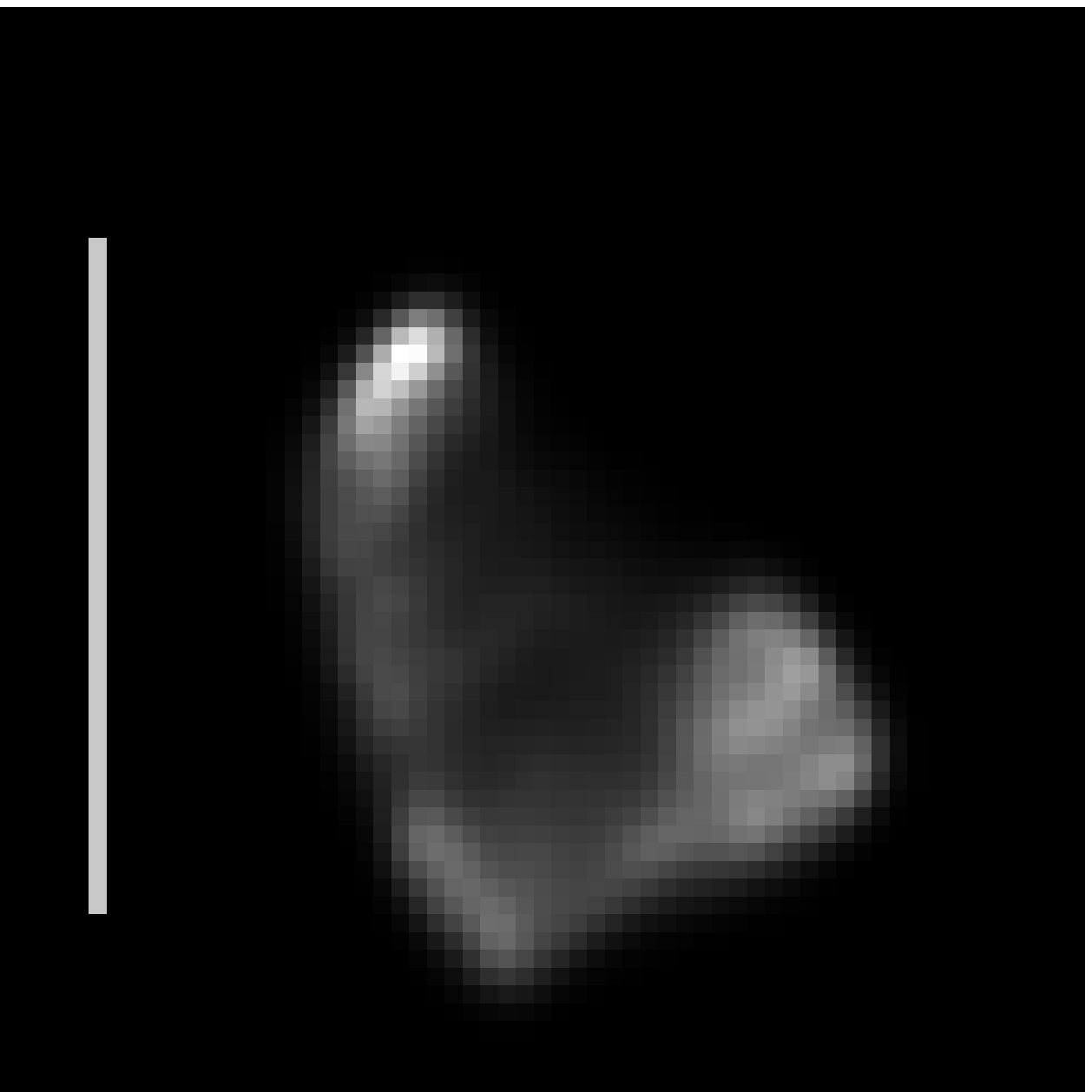}
  \epsfysize=2in \epsfbox{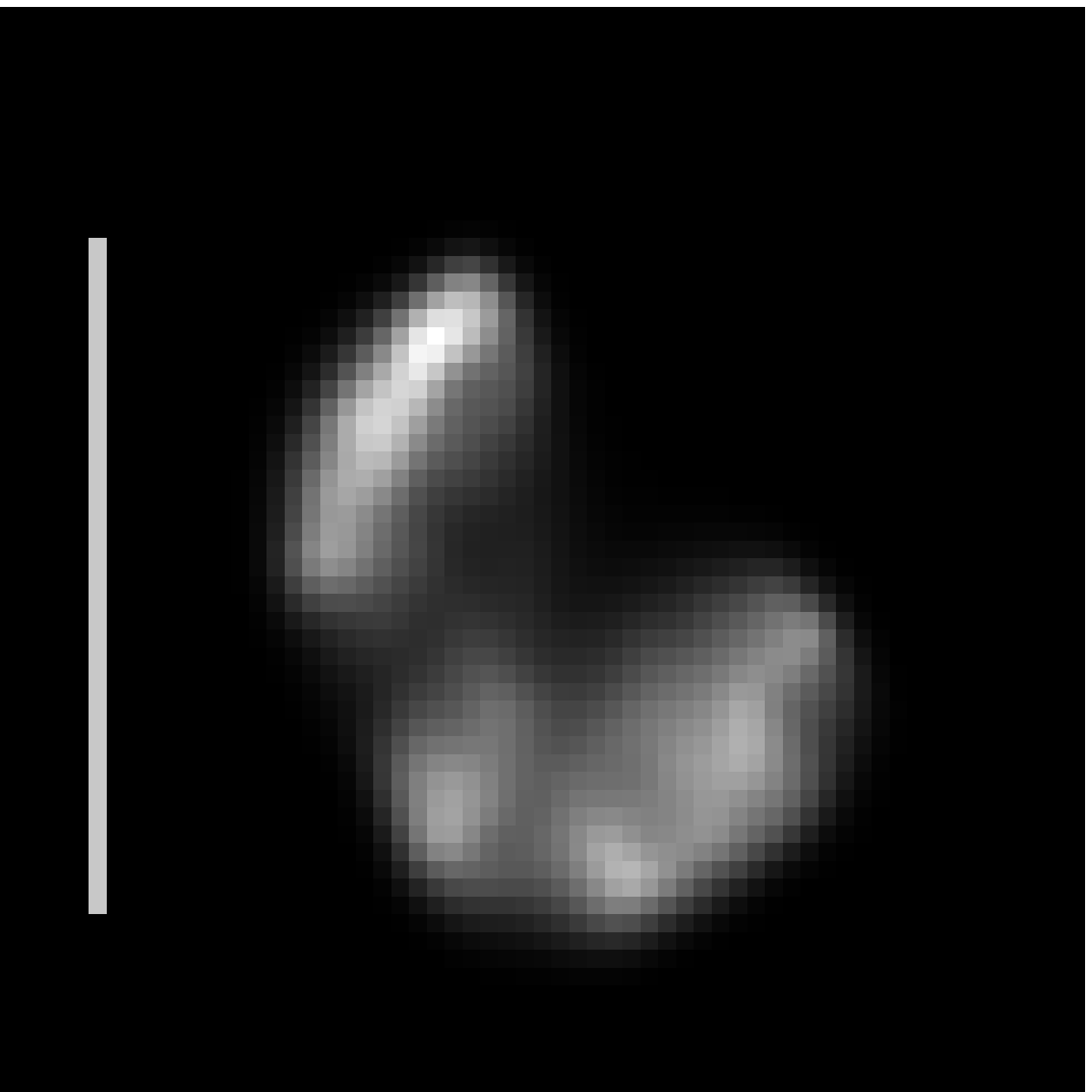}
} 
\centerline{
  \epsfysize=2in \epsfbox{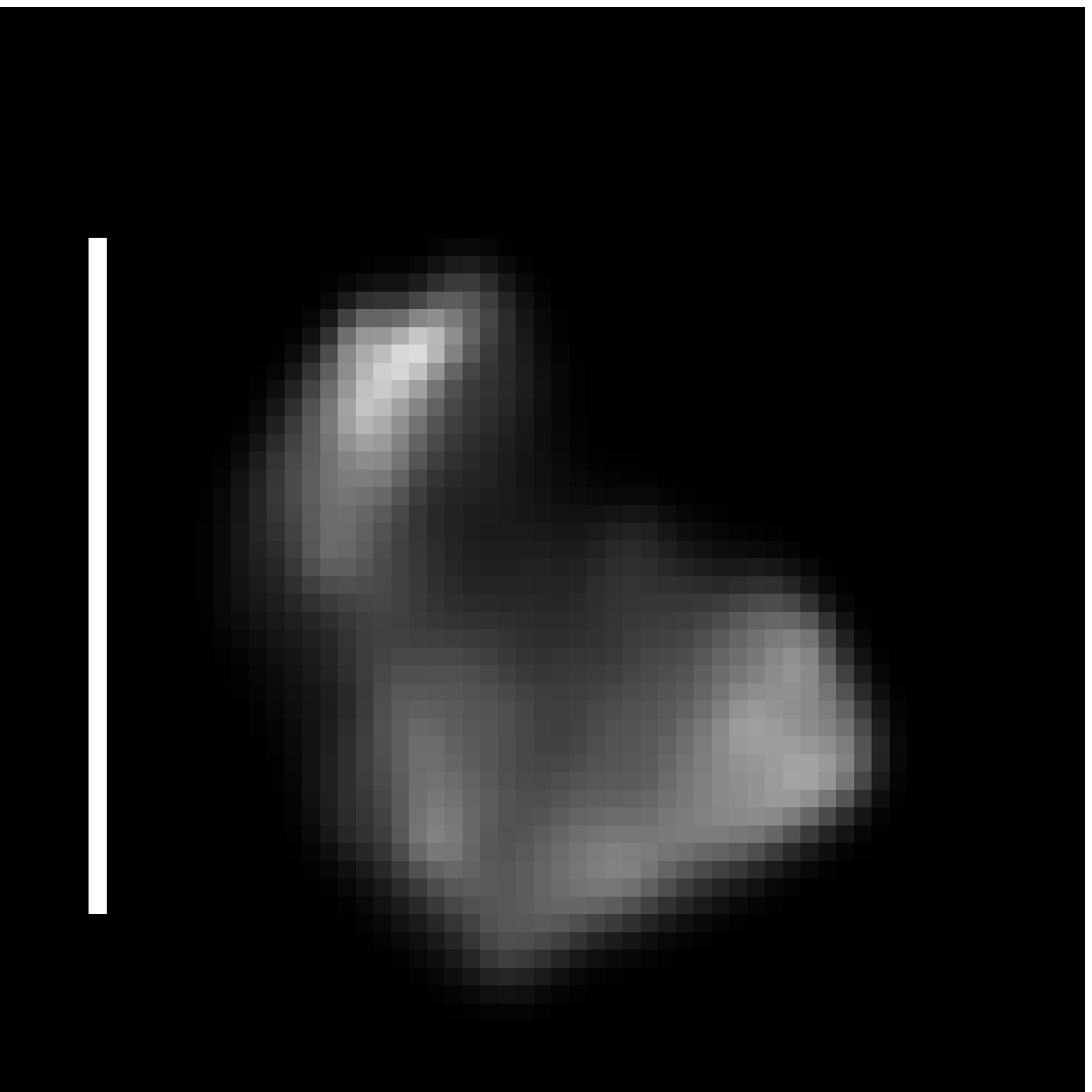}
} 
\end{figure}
Figure 2. Roberts et al.

\newpage

\begin{figure}[!ht]
\centerline{\epsfysize=4.0in \epsfbox{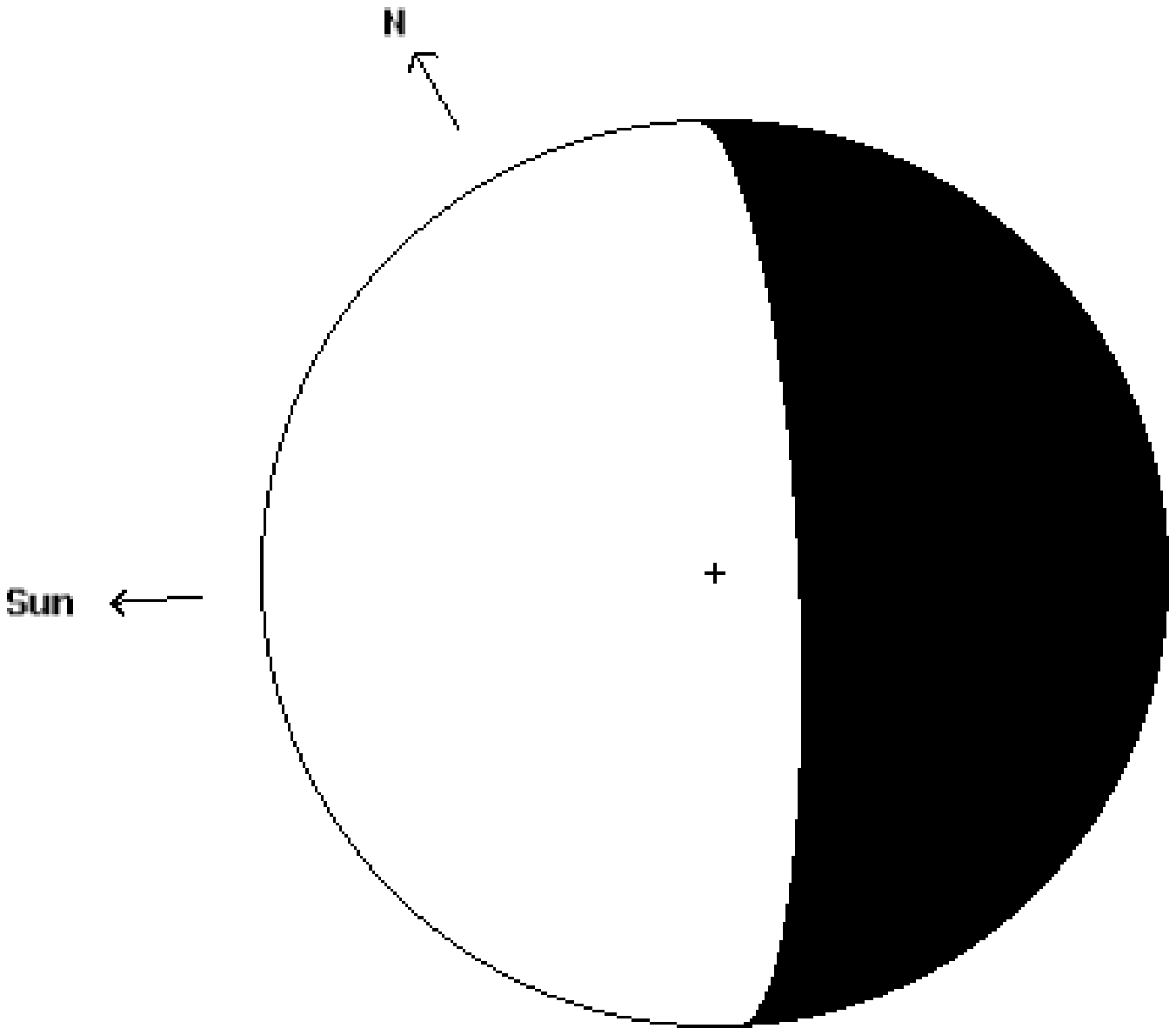}}
\end{figure}

Figure 3. Roberts et al. 

\newpage

\begin{figure}[!ht]
\centerline{\epsfysize=4.0in \epsfbox{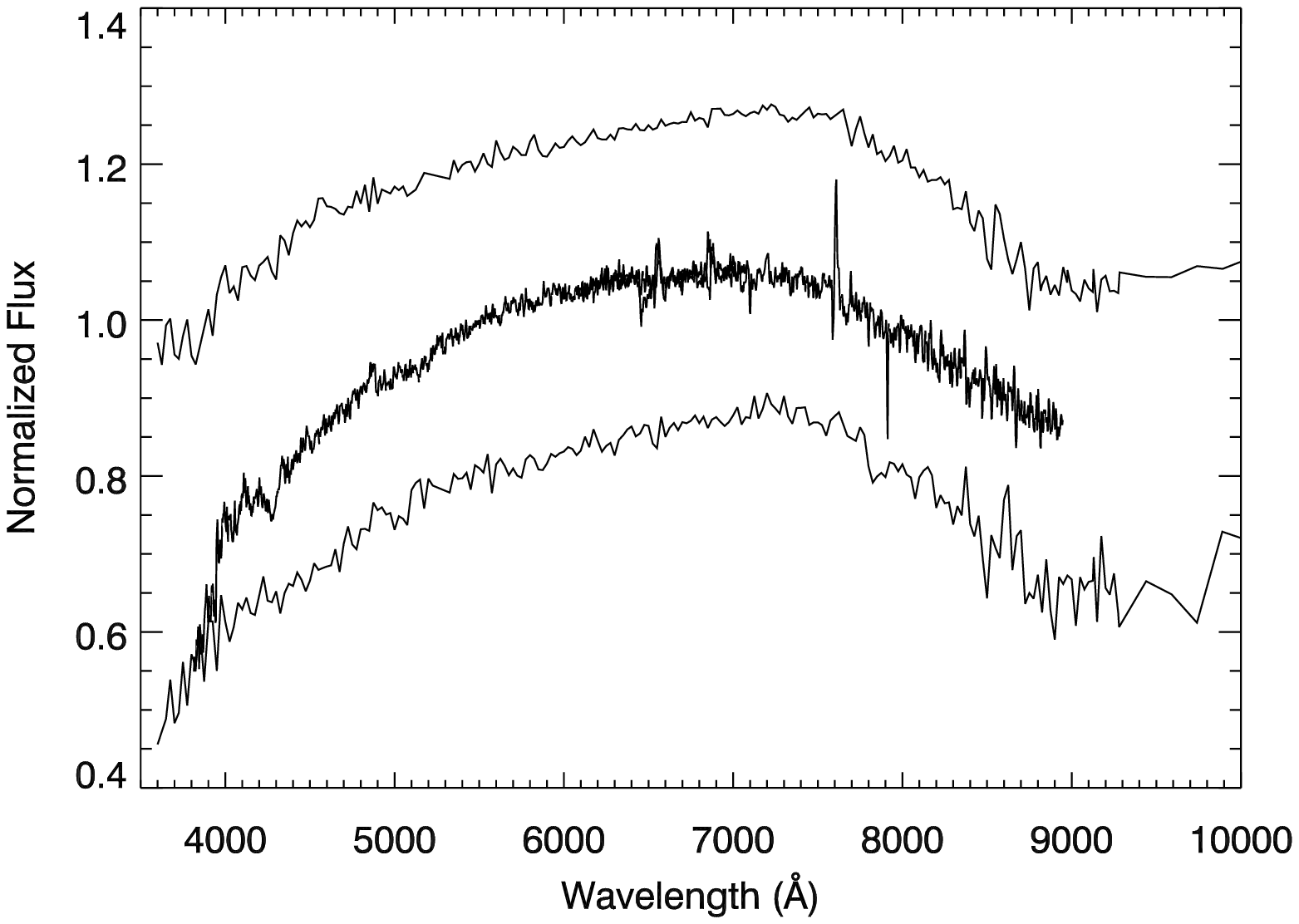}}
\end{figure}

Figure 4. Roberts et al.

\end{document}